\documentclass[USenglish,twocolumn]{article}
\usepackage[utf8]{inputenc}
\usepackage[big,online]{dgruyter}
\usepackage[sort&compress,square,numbers]{natbib}
\usepackage{times}

\begin{document}

  \DOI{10.1515/}
  \openaccess
  \pagenumbering{gobble}

\title{Automatic Generation of Synthetic Colonoscopy Videos for Domain Randomization}
\runningtitle{Automatic Generation of Synthetic Colonoscopy Videos for Domain Randomization}

\author[1]{Abhishek Dinkar Jagtap}
\author[1]{Mattias Heinrich}
\author*[1]{Marian Himstedt} 
\runningauthor{A.D.~Jagtap et al.}

\affil[1]{\protect\raggedright 
  Institute of Medical Informatics, University of Lübeck, Germany, e-mail: marian.himstedt@uni-luebeck.de}

\abstract{An increasing number of colonoscopic guidance and assistance systems rely on machine learning algorithms which require a large amount of high-quality training data. In order to ensure high performance, the latter has to resemble a substantial portion of possible configurations. This particularly addresses varying anatomy, mucosa appearance and image sensor characteristics which are likely deteriorated by motion blur and inadequate illumination. The limited amount of readily available training data hampers to account for all of these possible configurations which results in reduced generalization capabilities of machine learning models. We propose an exemplary solution for synthesizing colonoscopy videos with substantial appearance and anatomical variations which enables to learn discriminative domain-randomized representations of the interior colon while mimicking real-world settings. \footnote{Software components and video material will be made publicly available upon publication.} }


\maketitle

\section{Introduction} 
\thispagestyle{empty}
A common problem faced in machine learning is the lack of sufficient training data. For colonoscopy the majority of readily-available public image data is limited to individual frames or short sequences for benchmarking CAD-based polyp detection. Public colonoscopy videos of the entire colon structure are limited to rather low-quality capsule endoscopy video footage. The lack of ground truth camera poses further hampers the training of models for applications different from polyp detection such as: anatomical segment classification, visual place recognition (VPR), simultaneous localization and mapping (SLAM) and structure from motion (SfM). These applications require high-quality colonoscopy videos of the entire examinations covering all phases of the intervention.
\begin{center}
    \begin{figure}[ht]
    \centering
    \includegraphics[width=0.45\textwidth]{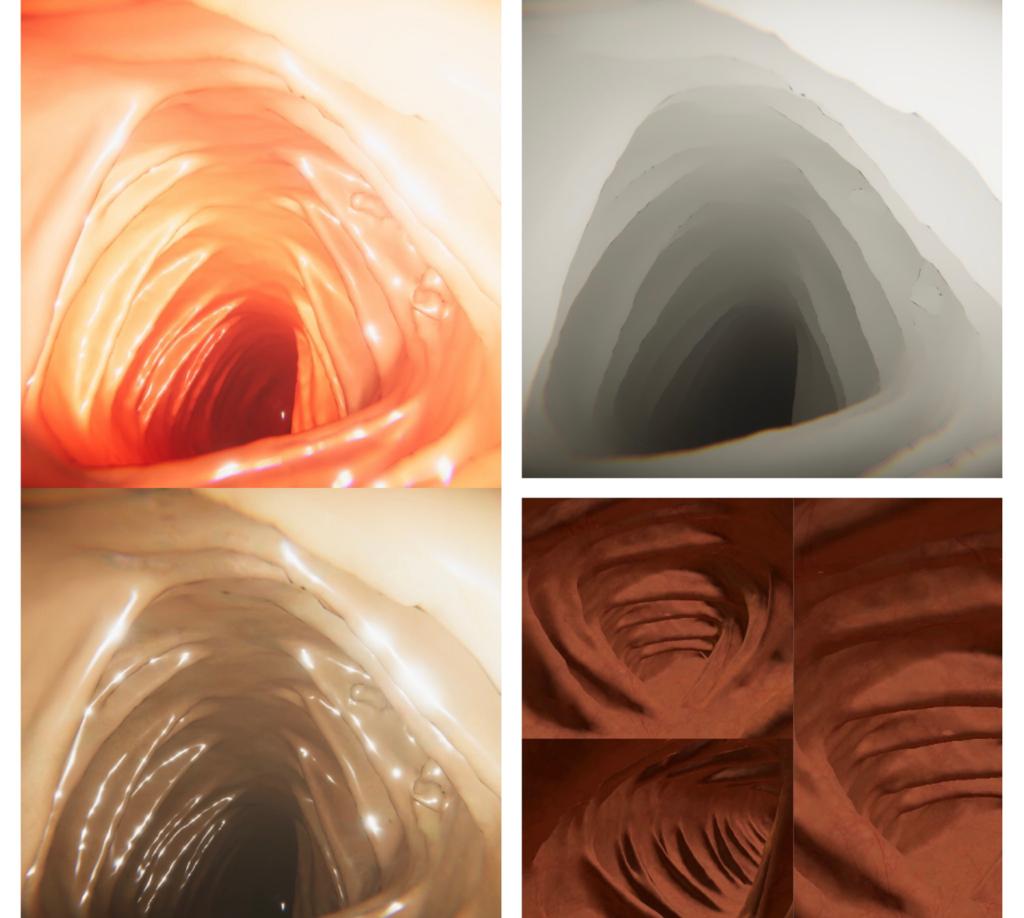}
    \caption{Examples of synthetic colonoscopy images.}
    \label{fig:synthetic_images}
\end{figure}
\end{center}

A common solution to this is the rendering of virtual endoscopy (VE) videos based on CT colonography data. VE provides both, image sequences and ground truth poses of varying anatomy, but (without further investigation) differs substantially from the visual appearance of real colonoscopy images. This entails gaps that have to be addressed by proper domain adaptation methods as demonstrated in \cite{mathew2020augmenting}. This, however, implies that synthesized images resemble colonoscopy images (and their anatomical locations) of small datasets which likely do not generalize well to unseen or less observed colon regions. 
Domain randomization, in contrast, utilizes a large amount of data which is randomly sampled over the entire configuration space with the variables being carefully predefined. It is important to note that domain randomization is practically applicable to only simulated data as some of the parameters such as textures, materials, occlusions and coat masks have to be properly controlled in a simulated environment have to be more elaborated than for generating VE images in order to enable visual appearance close to real colonoscopy images (see Fig. \ref{fig:synthetic_images}). Powerful engines such as \textit{Unity} have gained particular interest in the computer vision and robotics communities \cite{borkman2021unity,tremblay2018training}, but have been rarely investigated in medical imaging \cite{incetan2020vrcaps,billot2021synthseg}. 

\begin{figure*}[ht]
    \centering
    \includegraphics[width=0.8\textwidth]{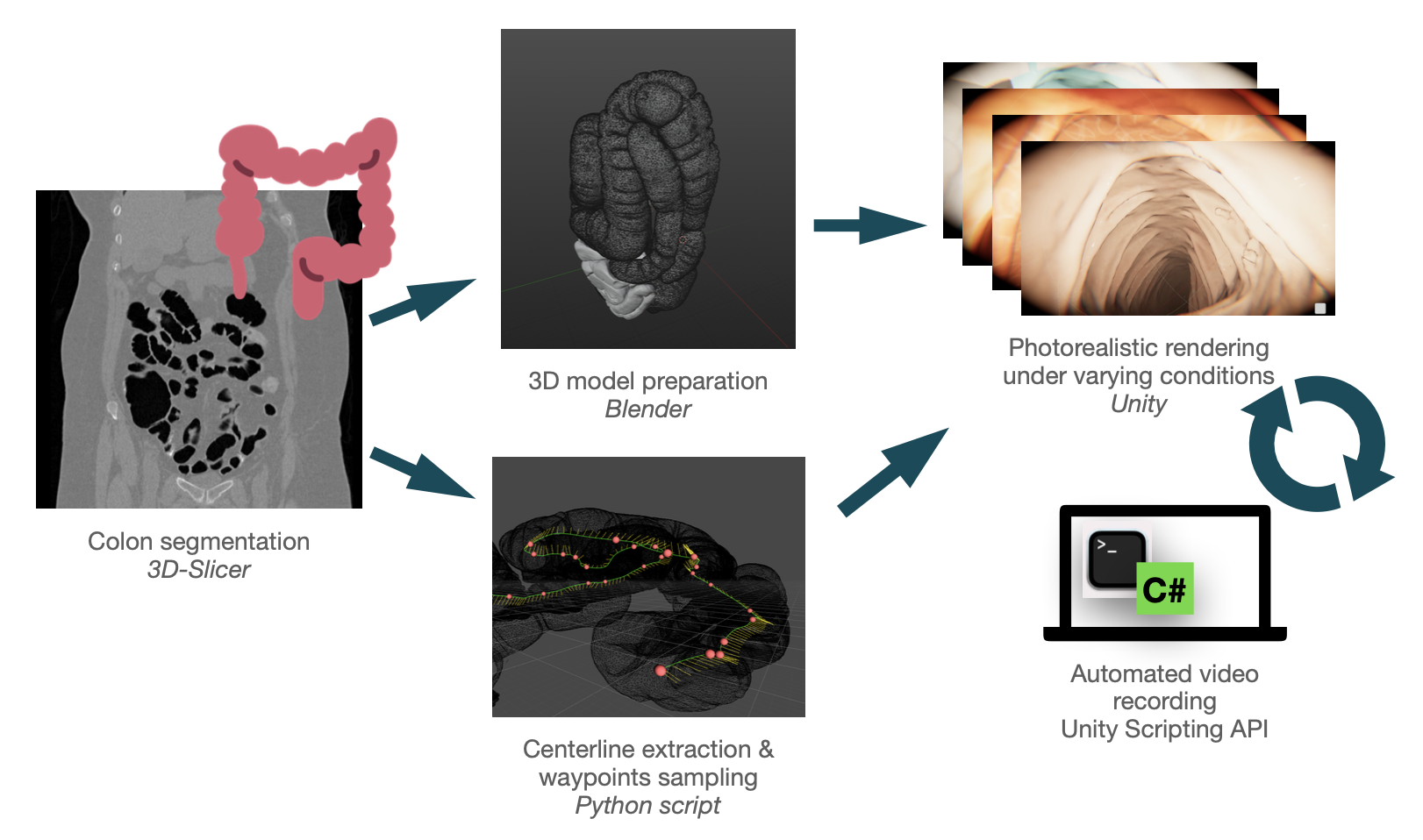}
    \caption{Overview of the utilized processing pipeline for generating synthetic images.}
    \label{fig:architecture}
\end{figure*}

Given sufficient capabilities of simulation, models can solely be trained on domain-randomized data while still achieving high generalization performance for inference on real-world test data.

This paper presents an exemplary implementation of domain randomization for colonoscopy with all required algorithmic components. It is built up on prior work \cite{incetan2020vrcaps} and supplements the latter by automated domain-randomized video recording through following waypoints along the interior colon's centerline.


\section{Material and Methods} 

\subsection{Colon segmentation}
At first, a CT colonography (CT with radiocontrast material) obtained from TCIA is imported in \textit{3D Slicer} for semi-automatic colon segmentation which is carried out as follows. A ROI around the colon is set manually with its image content being  thresholded. Subsequently we apply region-based segmentation on the (thresholded) mask to further delineate the colon structure. The segmentation mask is manually curated to ensure optimal results for successive steps.

\subsection{Centerline extraction}
For automated image collection we require an appropriate camera path through the interior colon. For this purpose, we estimate the centerline within the colon structure based on the prior work of \cite{wan2002automatic}. The key idea is to plan an obstacle-free (w.r.t colon wall) path from the anus (colon entry) to the caecum. Since the intuitive approach based on the shortest path estimation tends to get too close to corners in turns, Wan et al. propose to explicitly incorporate the inversed map of distances to the colon wall \cite{wan2002automatic} which was demonstrated to achieve optimal results with paths being exactly centered. Subsequently we sample equidistant waypoints along the extracted centerline which will be utilized within the simulation. Currently, we manually pick start and end points of the centerline extraction which, however, could be replaced by automatic anatomical landmark prediction through heatmap regression.

\subsection{3D model preparation}
Next, the colon segmentation is imported in \textit{Blender} for UV editing. Generally, a mesh is created surrounding the organ that can be edited along the vertices of the object. This mesh allows \textit{UV mapping} which is a method for projecting a 3D model surface onto a 2D plane for texture mapping. An UV editing tool as part of \textit{Blender} offers the possibility of unwrapping the 3D object onto a 2D plane where textures can be applied seamlessly throughout the region of the colon. This texture gives a realistic pattern to the object. Default shaders in \textit{Blender} enable to change material properties corresponding to colon such as surface IOR, secular tint and anisotropy to further enhance the realism. 

\subsection{Photorealistic rendering}
The 3D model prepared in \textit{Blender} is subsequently imported in \textit{Unity} which provides high definition render pipelines for our simulation environment that can produce photorealistic colonoscopy images. This virtual engine is commonly used for game development and has drawn particular interest in computer vision research due to its powerful graphical simulation platform for generating synthetic images. Using \textit{Unity} we are able to synthesize images where parameters such as lighting, materials, occlusions, transparency and coat mask are altered to give it a more realistic appearance. These parameters are carefully selected such that real-world characteristics are optimally mimicked.  As a starting base we utilize parts of the \textit{VR-Caps} project simulating a capsule endoscopic camera within \textit{Unity} \cite{incetan2020vrcaps}. A 3D model of this capsule with predefined attributes of an attached camera is placed inside the colon which is used for data collection. Adjusting these parameters is crucial for both mimicking real endoscopy and augmenting the data. The table below shows the camera parameters and post-processing effects required to achieve a fully synthetic model of the colon. For potential navigation tasks it is possible to additionally store corresponding depth images. 

\begin{table}[!h]
\label{tab:fonts}
\centering
\begin{tabular}{|c|c| }
\hline
Attributes &  Values\\
\hline
Surface Metallic 				& 0.3 \\
Surface Smoothness			& 0.7 \\
Lens Intensity 			& 0.1\\
Chromic Abberation			& 0.5 \\
Coat Mask			& 0.435	 \\
Camera's Field of View 			& 91.375	 \\
Focal length				& 159.45\\
ISO 		& 200 \\
Aperture 				& 16\\
Anisotropy 			& 1 \\
\hline
\end{tabular}
\caption{Camera parameters and Post-processing Effects}
\end{table}

\begin{figure*}
    \centering
    \includegraphics[width=\textwidth]{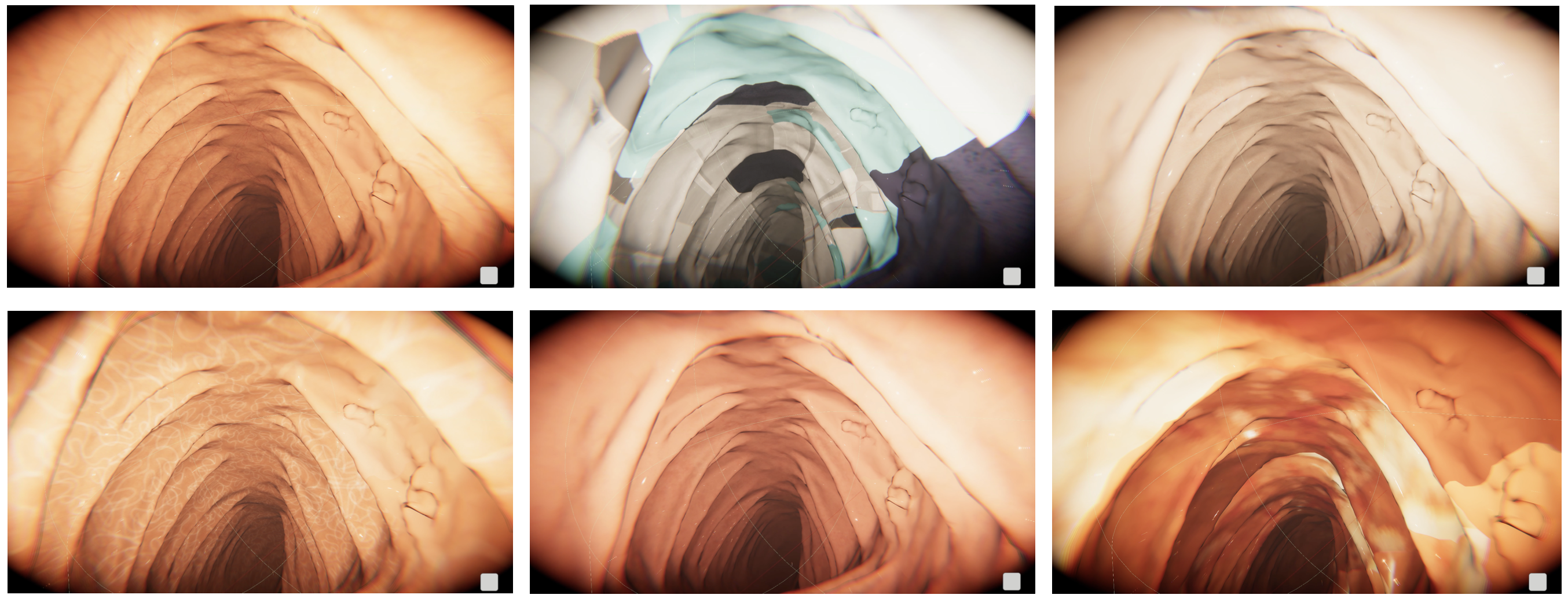}
    \caption{Synthesized, domain-randomized images captured at the same pose inside the colon. Textures are obtained as from random patterns as well as synthetic patterns mimicking mucosa appearances.}
    \label{fig:dr-samples}
\end{figure*}

\subsection{Automated Video Rendering}
Manually collecting data for endoscopy becomes highly time-consuming when creating synthetic datasets consisting of all the required variations and diversity. For domain randomization we need to record sequences of images each time with different textures and materials which entails substantial individual setup. Thus, an approach for automating the process of data collection is introduced, which allows us to collect numerous samples inside the colon with different parameters. For this purpose we make use of the \textit{scripting API} offered by \textit{Unity} which gives access to the simulation environment and interactive components via executable scripts. Firstly, the simulated capsule is introduced to the colon and then automatically steered along the waypoints of the centerline (see Fig. \ref{fig:waypoints}). The \textit{Unity engine} is setup appropriately such that it enables smooth camera motion while following the waypoints.  
Our path following script consist of two parts:  an \textit{initialization} function which runs all required initial setups (parameter setup, initial capsule pose) and an \textit{update} function which constantly controls the movement of the capsule (along the waypoints) and triggers actions such as changing parameters (e.g. lightning, texture). All images captured by the camera of the capsule are recorded. The parameters can either be adjusted on the fly allowing to capture images at the same pose with varying conditions or alternatively it is possible to alter the parameter set only for entire traversals. \textit{Unity} also enables to configure the capsule's speed and camera's field of view and targeted frame rate (FPS). 

\section{Results and Discussion} 

\begin{figure}
    \centering
    \includegraphics[width=0.5\textwidth]{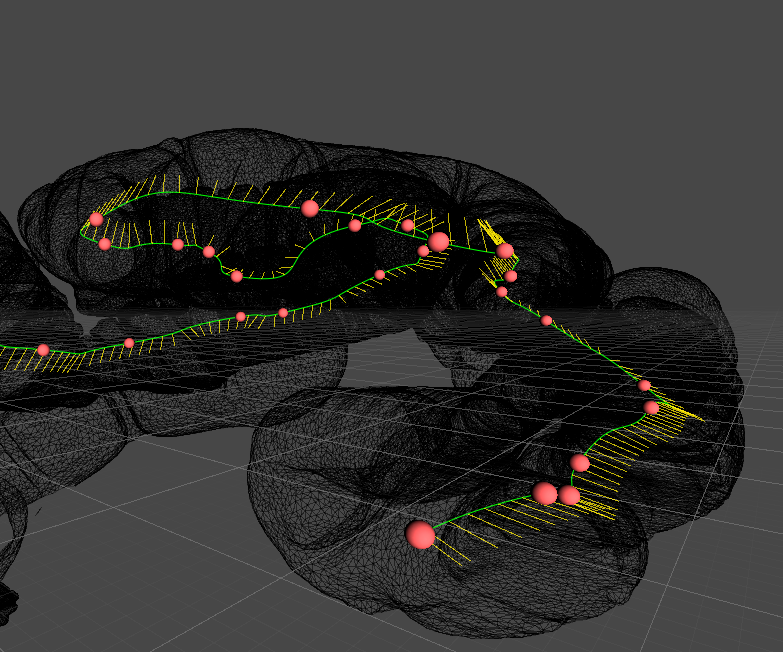}
    \caption{Path following the centerline of the colon. The green line visualize the path and red circles waypoints being traced by the simulated capsule.}
    \label{fig:waypoints}
\end{figure}

\begin{figure}
    \centering
    \includegraphics[width=0.5\textwidth]{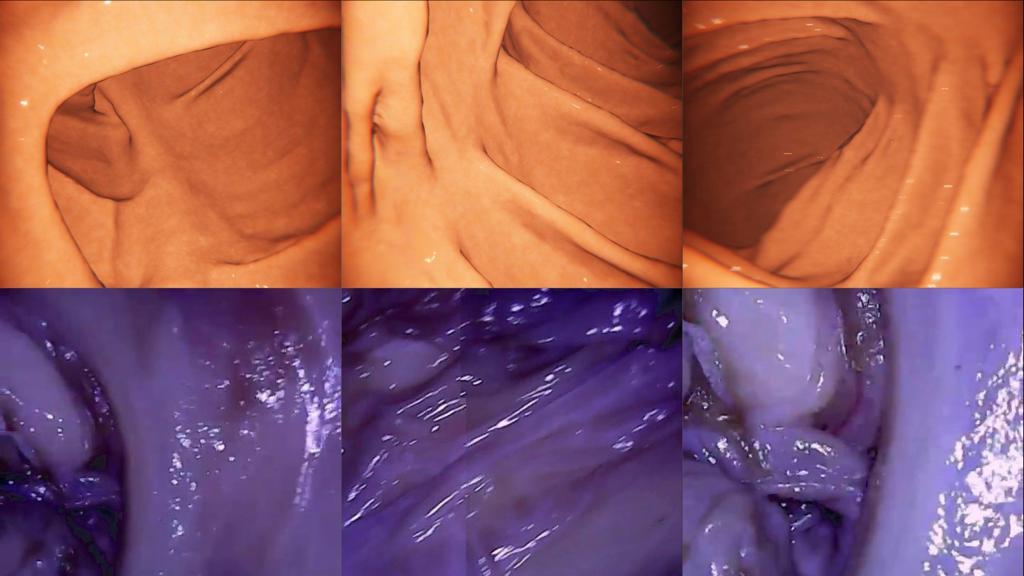}
    \caption{Comparison of synthetic (top) and real (bottom) colonoscopy images.}
    \label{fig:real_samples}
\end{figure}

We evaluate our simulation qualitatively based on image renderings for varying parameters which becomes particularly apparent when randomizing surface material and textural patterns. This is illustrated by Fig.\ref{fig:dr-samples} which shows different renderings from the same captured from the same camera pose inside the colon. 
Fig. \ref{fig:waypoints} shows an example of an extracted centerline and generated waypoints being followed for automated video recording. For comparison, Fig. \ref{fig:real_samples} shows exemplary real and synthetic images respectively.

\section{Conclusion}
This paper presented a pipeline for generating synthetic colonoscopy videos that can be used to improve training of deep learning models. By controlling environment (e.g. texture, reflectance) as well as virtual camera (e.g. lightning) properties we are able to simulate conditions being observed in inference but hardly ever presented in training data which is particularly the case for small-scale datasets. Inspired by substantive improvements reported on computer vision and robotics applications and limited prior work (VR-Caps)  \cite{incetan2020vrcaps} we motivate to utilize domain-randomized synthesization for video colonoscopy. In our future work, we will incorporate this additional data for training deep learning-based approaches to SfM, SLAM and 3D reconstruction. In order to further simplify the variation in patient anatomy, we will investigate (fully) automatic segmentation of the colon in CT scans as well as an alternative to the 3D model preparation in \textit{Blender}. 
\pagebreak

\bibliographystyle{IEEEtran}
%
\bibliography{mybiblefile}

\begin{thebibliography}{1}
\providecommand{\url}[1]{#1}
\csname url@samestyle\endcsname
\providecommand{\newblock}{\relax}
\providecommand{\bibinfo}[2]{#2}
\providecommand{\BIBentrySTDinterwordspacing}{\spaceskip=0pt\relax}
\providecommand{\BIBentryALTinterwordstretchfactor}{4}
\providecommand{\BIBentryALTinterwordspacing}{\spaceskip=\fontdimen2\font plus
\BIBentryALTinterwordstretchfactor\fontdimen3\font minus
  \fontdimen4\font\relax}
\providecommand{\BIBforeignlanguage}[2]{{%
\expandafter\ifx\csname l@#1\endcsname\relax
\typeout{** WARNING: IEEEtran.bst: No hyphenation pattern has been}%
\typeout{** loaded for the language `#1'. Using the pattern for}%
\typeout{** the default language instead.}%
\else
\language=\csname l@#1\endcsname
\fi
#2}}
\providecommand{\BIBdecl}{\relax}
\BIBdecl

\bibitem{borkman2021unity}
S.~Borkman, A.~Crespi, S.~Dhakad, S.~Ganguly, J.~Hogins, Y.-C. Jhang,
  M.~Kamalzadeh, B.~Li, S.~Leal, P.~Parisi \emph{et~al.}, ``Unity perception:
  Generate synthetic data for computer vision,'' \emph{arXiv preprint
  arXiv:2107.04259}, 2021.

\bibitem{incetan2020vrcaps}
K.~Incetan, I.~O. Celik, A.~Obeid, G.~I. Gokceler, K.~B. Ozyoruk,
  Y.~Almalioglu, R.~J. Chen, F.~Mahmood, H.~Gilbert, N.~J. Durr, and M.~Turan,
  ``Vr-caps: A virtual environment for capsule endoscopy,'' 2020.

\bibitem{choudhary2020advancing}
A.~Choudhary, L.~Tong, Y.~Zhu, and M.~D. Wang, ``Advancing medical imaging
  informatics by deep learning-based domain adaptation,'' \emph{Yearbook of
  medical informatics}, vol.~29, no.~01, pp. 129--138, 2020.

\bibitem{juliani2018unity}
A.~Juliani, V.-P. Berges, E.~Teng, A.~Cohen, J.~Harper, C.~Elion, C.~Goy,
  Y.~Gao, H.~Henry, M.~Mattar \emph{et~al.}, ``Unity: A general platform for
  intelligent agents,'' \emph{arXiv preprint arXiv:1809.02627}, 2018.

\bibitem{reynisson2015airway}
P.~J. Reynisson, M.~Scali, E.~Smistad, E.~F. Hofstad, H.~O. Leira, F.~Lindseth,
  T.~A. Nagelhus~Hernes, T.~Amundsen, H.~Sorger, and T.~Lang{\o}, ``Airway
  segmentation and centerline extraction from thoracic ct--comparison of a new
  method to state of the art commercialized methods,'' \emph{PloS one},
  vol.~10, no.~12, p. e0144282, 2015.

\bibitem{wan2002automatic}
M.~Wan, Z.~Liang, Q.~Ke, L.~Hong, I.~Bitter, and A.~Kaufman, ``Automatic
  centerline extraction for virtual colonoscopy,'' \emph{IEEE transactions on
  medical imaging}, vol.~21, no.~12, pp. 1450--1460, 2002.

\bibitem{nikolenko2021synthetic}
S.~I. Nikolenko, ``Synthetic data for deep learning,'' 2021.

\bibitem{seib2020mixing}
V.~Seib, B.~Lange, and S.~Wirtz, ``Mixing real and synthetic data to enhance
  neural network training--a review of current approaches,'' \emph{arXiv
  preprint arXiv:2007.08781}, 2020.

\end{thebibliography}

\end{document}